\documentclass[pra,twocolumn,floatfix,a4paper,superscriptaddress]{revtex4}

\usepackage{bm,color,amsmath,txfonts}
\usepackage{graphicx}
\usepackage{siunitx}
\usepackage{subfigure}
\usepackage{verbatim}
\usepackage{dcolumn}
\usepackage{bm}
\usepackage{epsf}
\usepackage{xcolor}
\usepackage{hyperref}
\usepackage{hhline}
\usepackage{float}
\usepackage{enumerate}
\usepackage{bbm}
\usepackage{lipsum}

\begin{document}
\title{Entangling mechanical vibrations of two massive ferrimagnets by fully exploiting \\ the nonlinearity of magnetostriction}
\author{Hang Qian}
\affiliation{Interdisciplinary Center of Quantum Information, State Key Laboratory of Modern Optical Instrumentation, and Zhejiang Province Key Laboratory of Quantum Technology and Device, Department of Physics, Zhejiang University, Hangzhou 310027, China}
\author{Zhi-Yuan Fan}
\affiliation{Interdisciplinary Center of Quantum Information, State Key Laboratory of Modern Optical Instrumentation, and Zhejiang Province Key Laboratory of Quantum Technology and Device, Department of Physics, Zhejiang University, Hangzhou 310027, China}
\author{Jie Li}\thanks{jieli007@zju.edu.cn}
\affiliation{Interdisciplinary Center of Quantum Information, State Key Laboratory of Modern Optical Instrumentation, and Zhejiang Province Key Laboratory of Quantum Technology and Device, Department of Physics, Zhejiang University, Hangzhou 310027, China}

\begin{abstract}
Quantum entanglement in the motion of macroscopic objects is of significance to both fundamental studies and quantum technologies. Here we show how to entangle the mechanical vibration modes of two massive ferrimagnets that are placed in the same microwave cavity. Each ferrimagnet supports a magnon mode and a low-frequency vibration mode coupled by the magnetostrictive force. The two magnon modes are, respectively, coupled to the microwave cavity by the magnetic dipole interaction. We first generate a stationary nonlocal entangled state between the vibration mode of the ferrimagnet-1 and the magnon mode of the ferrimagnet-2. This is realized by continuously driving the ferrimagnet-1 with a strong red-detuned microwave field and the entanglement is achieved by exploiting the magnomechanical parametric down-conversion and the cavity-magnon state-swap interaction. We then switch off the pump on the ferrimagnet-1 and, simultaneously, turn on a red-detuned pulsed drive on the ferrimagnet-2. The latter drive is used to activate the magnomechanical beamsplitter interaction, which swaps the magnonic and mechanical states of the ferrimagnet-2. Consequently, the previously generated phonon-magnon entanglement is transferred to the mechanical modes of two ferrimagnets. The work provides a scheme to prepare entangled states of mechanical motion of two massive objects, which may find applications in various studies exploiting macroscopic entangled states.
\end{abstract}

\date{\today}
\maketitle

\section{Introduction}
Entanglement of mechanical motion has been first demonstrated in two microscopic trapped atomic ions~\cite{Wineland}, then in two single-phonon excitations in nanodiamonds~\cite{Ian}, and recently in two macroscopic optomechanical resonators~\cite{SG,Mika,John}. In the past two decades, a number of theoretical proposals have been given in optomechanics~\cite{DV02,Peng,DV05,DV06,DV07,Hartmann,GSA09,Girvin,DV12,Clerk13,Tan,Savona,Liao,RX14,Clerk14,Abdi15,Jie15,Buchmann,Zip15,An,Ferraro,Zip16,Lv16,Jie17,Reid17,Sarma,L1,L2,L3,L4} for preparing the entanglement between massive mechanical resonators, utilizing the coupling between optical and mechanical degrees of freedom by radiation pressure. In particular, the application of reservoir engineering ideas~\cite{Zoller96,LD,Zoller08,Cirac,Morigi} to optomechanics~\cite{Clerk13,Tan,Clerk14,Jie15} has led to a significant and robust mechanical entanglement and the entanglement has been successfully demonstrated in the experiment~\cite{Mika}. 

Exploring novel physical platforms that could prepare quantum states at a more massive scale is of great significance for the study of macroscopic quantum phenomena~\cite{Gisin}, the boundary between the quantum and classical worlds~\cite{Bassi}, and gravitational quantum physics~\cite{Bose,VV}, etc. Recently, the magnomechanical system of a large-size ferrimagnet, e.g., yttrium iron garnet (YIG), that has a dispersive magnon-phonon coupling has shown such a potential~\cite{Jie18,Jie19a,Jie19,Tan2,CLi,Yang,Jie21,CLi2,HFW,Busch,JJ,JHW,HJ,HF2}. The magnon and mechanical vibration modes are coupled by the nonlinear magnetostrictive interaction, which couples ferrimagnetic magnons to the deformation displacement of the ferrimagnet~\cite{Kittel,Tang16,Davis,Jie22}. The magnomechanical Hamiltonian takes the same form as the optomechanical one~\cite{omRMP} (by exchanging the roles of magnons and photons), which allows us to predict many optomechanical analogues in magnomechanics. By coupling the magnomechanical system to a microwave cavity, they form the tripartite cavity magnomechanical system.  The magnetostriction has been exploited in cavity magnomechanics to generate macroscopic entangled states of magnons and vibration phonons of massive ferrimagnets~\cite{Jie18,Jie19,Tan2,Jie21,CLi2,JHW,HF2}, as well as nonclassical states of microwave fields~\cite{Jie20,NSR}.  Therefore, the magnetostrictive nonlinearity becomes a valuable resource for producing various quantum states of microwave photons, magnons, and phonons. These nonclassical states may find potential applications in quantum information processing~\cite{Naka19,Yuan}, quantum metrology~\cite{NSR} and quantum networks~\cite{prxQ}. Despite of the aforementioned many proposals and very limited experiments~\cite{Tang16,Davis,Jie22} in cavity magnomechanics, there is so far only one protocol~\cite{Jie21} for entangling two mechanical vibration modes of macroscopic ferrimagnets. The protocol~\cite{Jie21}, however, relies on an external entangled resource and transfers the quantum correlation from microwave drive fields to two vibration modes. Therefore, designing more energy-saving protocols without using any external quantum resource is highly needed.  

Along this line, we present here a scheme for generating a nonlocal entangled state between the mechanical vibrations of two ferrimagnets, without the need of any quantum driving field. Two ferrimagnets are placed in a microwave cavity and each ferrimagnet supports a magnon mode and a mechanical mode. The cavity mode couples to two magnon modes by the magnetic dipole interaction, and the magnon modes couple to their local vibration modes by magnetostriction, respectively. We show that the entanglement between the vibration modes of two ferrimagnets can be achieved by fully exploiting the nonlinear magnetostriction interaction (i.e., exploiting both the magnomechanical parametric down-conversion (PDC) and state-swap interactions) and by using the common cavity field being an intermediary to distribute quantum correlations. The mechanical entanglement is established by two steps. We first generate steady-state entanglement between the mechanical mode of the ferrimagnet-1 and the magnon mode of the ferrimagnet-2. We then activate the magnomechanical state-swap interaction in the ferrimagnet-2, which transfers the magnonic state to its locally interacting mechanical mode. Consequently, the two mechanical modes get nonlocally entangled.   

The remainder of the paper is organized as follows. In Sec.~\ref{model}, we introduce the general model of the protocol that is used in the two steps. We then show how to prepare a stationary nonlocal magnon-phonon entanglement with a continuous microwave pump in Sec.~\ref{1step}, and how to transfer this entanglement to two mechanical modes with a pulsed microwave drive in Sec.~\ref{2step}. Finally, we discuss and conclude in Sec.~\ref{conc}.

\section{The model}\label{model}

\begin{figure}[t]
\hskip-0.35cm \includegraphics[width=0.97\linewidth]{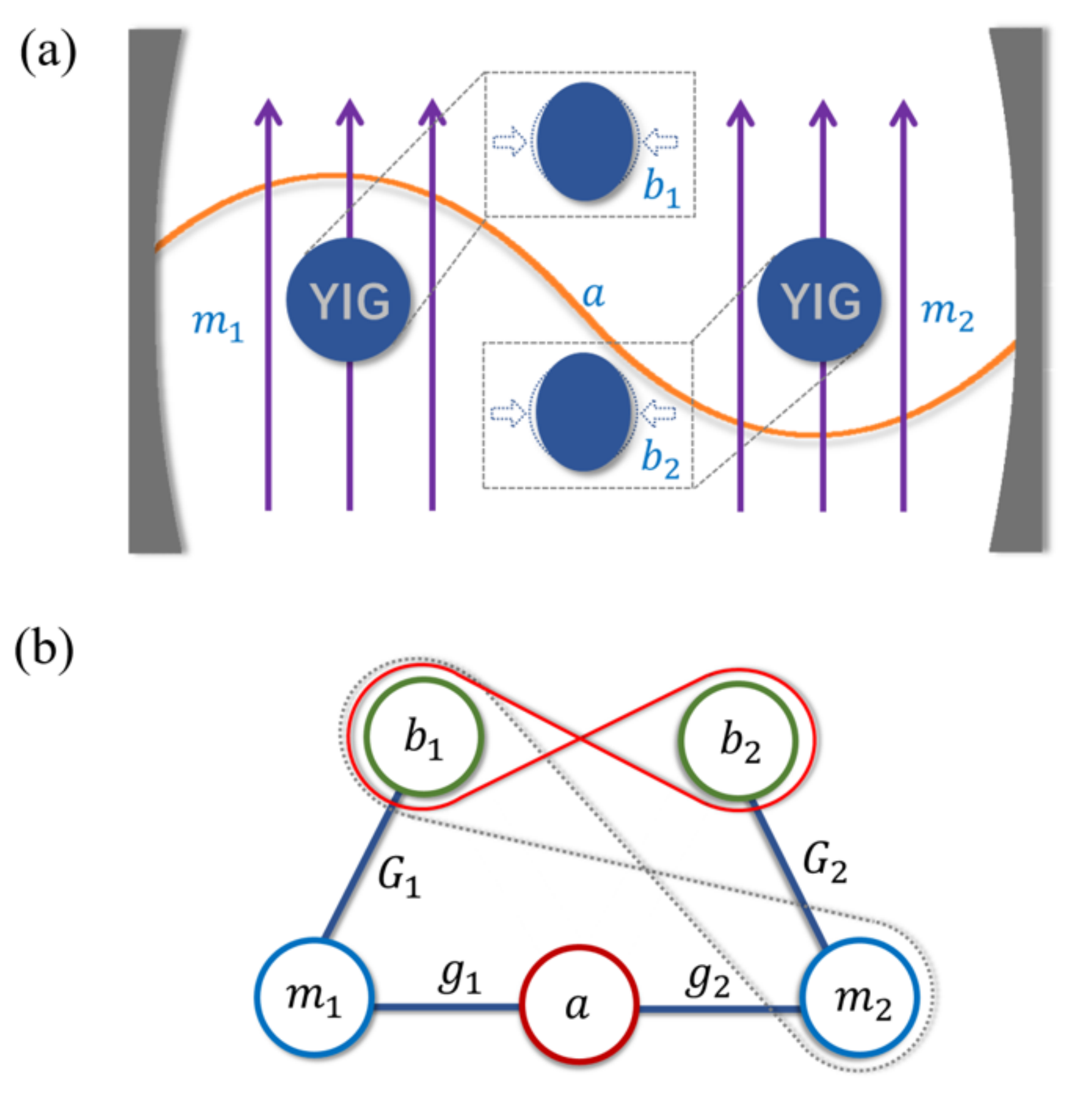}
\caption{(a) Sketch of the system. Two YIG crystals are placed near the maximum magnetic fields of a microwave cavity. Each YIG crystal is in a uniform bias magnetic field, and supports a magnon mode and a mechanical vibration mode. The two magnon modes couple to the same cavity field. Two drive fields are applied {\it successively} to the two magnon modes in the two steps of the protocol. Note that though YIG spheres are adopted in the sketch, the YIG crystals can be a nonspherical structure, e.g., micron-sized YIG bridges~\cite{bridge,Fan2}.  (b) Interactions among the subsystems. The magnon mode $m_j$ ($j=1,2$) couples linearly to the cavity mode $a$ with the coupling strength $g_j$, and couples dispersively to the mechanical mode $b_j$ with the effective magnomechanical coupling rate $G_j$. A nonlocal phonon-magnon ($b_1$-$m_2$) entanglement is created in the first step with a continuous drive on the magnon mode $m_1$, and the entanglement is transferred to the two mechanical modes in the second step by using a pulsed drive on the magnon mode $m_2$.  }
\label{fig1}
\end{figure}

The protocol is based on a hybrid five-mode cavity magnomechanical system, including a microwave cavity mode, two magnon modes, and two mechanical vibration modes, as depicted in Fig.~\ref{fig1}. The magnon modes are embodied by the collective motion of a large number of spins (i.e., spin wave) in two ferrimagnets, e.g., two YIG spheres~\cite{Tang16,Davis,Jie22} or micro bridges~\cite{bridge,Fan2}. They simultaneously couple to the microwave cavity via the magnetic dipole interaction. This coupling can be strong thanks to the high spin density in YIG~\cite{S1,S2,S3}. The mechanical modes refer to the deformation vibration modes of two YIG crystals caused by the magnetostrictive force. Due to the much lower mechanical frequency (ranging from $10^0$ to $10^2$ MHz) than the magnon frequency (GHz) in the typical magnomechanical systems~\cite{Tang16,Davis,Jie22,bridge}, the vibration phonons and the magnons are coupled in a {\it dispersive} manner~\cite{Fan2,Fan,Oriol}. The Hamiltonian of the system reads
\begin{equation}\label{Hsys}
\begin{split}
  H/\hbar = \, &\omega_{a} a^{\dagger}a+\sum_{j=1,2}\left( \omega_{m_j}m_{j}^{\dagger}m_{j}+\frac{\omega_{b_j}}{2}\left(p_j^2+q_j^2\right) \right) \\
  &+\sum_{j=1,2} \bigg( g_j\left(am_j^\dagger+a^\dagger m_j\right) + G_{0j}m_j^\dagger m_j q_j  \bigg) \\ 
  &+i\Omega_k\left(m_k^\dagger e^{-i\omega_{0k} t}-m_k e^{i\omega_{0k} t}\right).
\end{split}
\end{equation}
The first (second) term describes the energy of the cavity mode (magnon modes), of which the frequency is $\omega_a$ ($\omega_{m_j}$) and the annihilation operator is $a$ ($m_j$) with the commutation relation $[a,a^\dagger]=1$ $\left(\big[m_j,m_j^\dagger \big]=1\right)$. The magnon frequency $\omega_{m_j}$ is determined by the external bias magnetic field $H_j$ via $\omega_{m_j}=\gamma H_j$, where the gyromagnetic ratio $\gamma/2\pi = \SI{28}{GHz/T}$. The third term denotes the energy of two mechanical vibration modes with frequencies $\omega_{b_j}$, and $q_j$ and $p_j$ ($[q_j,p_j ]=i$) are the dimensionless position and momentum of the vibration mode $j$, modeled as a mechanical oscillator. The coupling $g_{j}$ is the linear cavity-magnon coupling rate, and $G_{0j}$ is the bare magnomechanical coupling rate. For large-size YIG spheres with the diameter in the 100 $\mu$m range~\cite{Tang16,Davis,Jie22}, $G_{0j}$ is typically in the 10 mHz range~\cite{Oriol}. It, however, can be much stronger for micron-sized YIG bridges~\cite{bridge,Fan2}. Nevertheless, the effective magnomechanical coupling can be significantly enhanced by driving the magnon mode with a strong microwave field~\cite{Jie18}. The driving Hamiltonian is described by the last term, and the corresponding Rabi frequency $\Omega_{k}=\frac{\sqrt{5}}{4}\gamma\sqrt{N_k}B_{k}$ ($k=1$ or 2)~\cite{Jie18}, with $B_{k}$ ($\omega_{0k}$) being the amplitude (frequency) of the drive magnetic field, and $N_k$ being the total number of spins in the $k$th crystal. 
We remark that the model differs from the one used in Ref.~\cite{Jie19} by including a second mechanical mode, which brings in a significant amount of additional thermal noise to the system. More importantly, the present work aims to entangle two low-frequency (in MHz) mechanical modes. This is much more difficult to prepare than the entanglement of two GHz magnon modes studied in Ref.~\cite{Jie19}.

In what follows, we adopt a two-step procedure to prepare the two mechanical modes in an entangled state, and in each step, we apply a single drive field on either magnon mode $m_1$ or $m_2$. This avoids the complex Floquet dynamics in our highly hybrid system caused by simultaneously applying multiple pump tones~\cite{Clerk13,Tan,Clerk14,Jie15}. We first generate a nonlocal entangled state between the mechanical mode $b_1$ and the magnon mode $m_2$ by continuously driving the magnon mode $m_1$. After the system enters a stationary state, we then turn off the drive on $m_1$ and, simultaneously, turn on a red-detuned drive on the magnon mode $m_2$ to activate the magnomechanical state-swap interaction $m_2 \leftrightarrow b_2$. This operation transfers the quantum correlation shared between $b_1$ and $m_2$ to two mechanical modes, thus establishing a quantum correlation (i.e., entanglement) between the two mechanical modes.

\begin{figure}
\hskip-0.0cm \includegraphics[width=\linewidth]{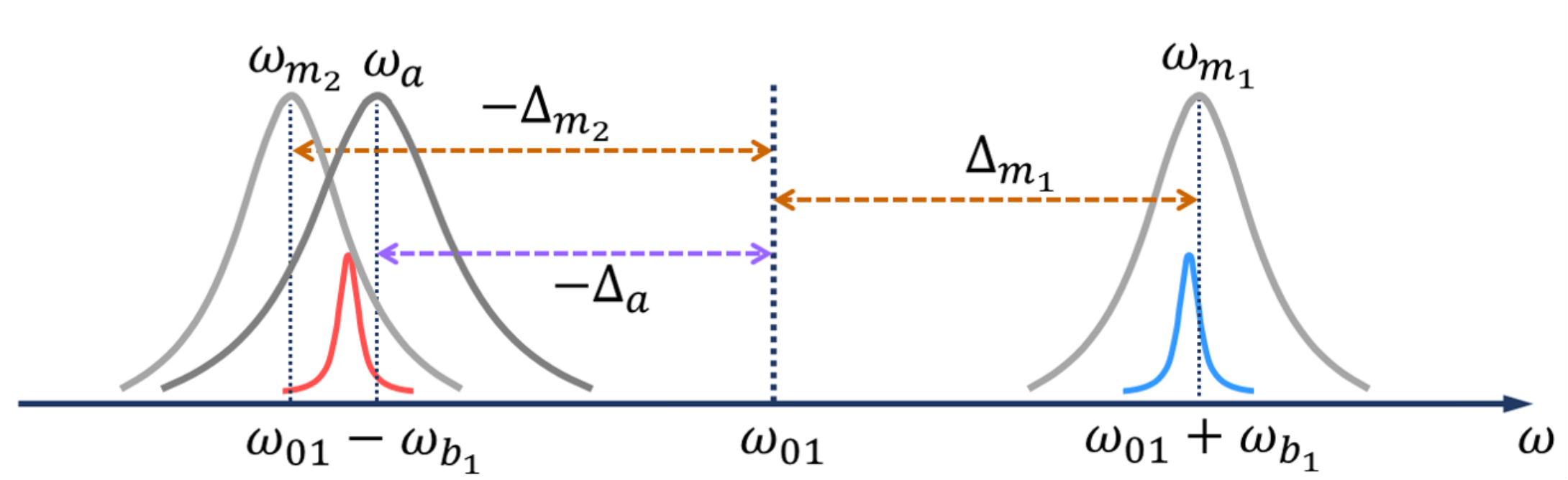}
\caption{Mode and drive frequencies used in the first step. When the magnon mode $m_1$ is (the cavity and magnon mode $m_2$ are) resonant with the blue (red) mechanical sideband of the drive field with frequency $\omega_{01}$, the nonlocal phonon-magnon entanglement $E_{b_1 m_2}$ is established.}
\label{fig2}
\end{figure}

\section{Stationary nonlocal magno-mechanical entanglement}
\label{1step}

In the first step, we aim to entangle the mechanical mode $b_1$ and the magnon mode $m_2$. This can be realized by driving the magnon mode $m_1$ with a {\it strong} red-detuned microwave field~\cite{Jie18}, see Fig.~\ref{fig2}. It was shown that a genuine tripartite magnon-photon-phonon entangled state can be produced without involving the second YIG crystal~\cite{Jie18}. By using the partial result that the mechanical mode $b_1$ and the cavity $a$ are entangled, and by coupling the cavity to the second {\it nearly resonant} magnon mode $m_2$ (which have a beamsplitter interaction realizing the state-swap operation $a \leftrightarrow m_2$), the two modes $b_1$ and $m_2$ thus get entangled. This is confirmed by the numerical results presented in this section.

It should be noted that since the strong drive is applied on the {\it first} YIG crystal, the effective (magnomechanical) coupling to the mechanical mode $b_2$ in the {\it second} YIG crystal is much smaller than that in the first YIG crystal, $G_2 \ll G_1$, so the presence of the second mechanical mode, or not, will not appreciably affect the entanglement dynamics analysed above. Because in the next step, the coupling to the second mechanical mode must be turned on ($G_{02}>0$), including this coupling in the model also in the first step means no additional operation (e.g., adjusting the direction of the bias magnetic field to activate or inactivate the coupling $G_{02}$~\cite{Tang16,Jie19}) has to be implemented between the two steps. Another reason is that, as will be shown in Sec.~\ref{2step}, the entanglement between $b_1$-$m_2$ should be transferred to the two mechanical modes as soon as possible because it rapidly decays when the drive in the first step is switched off.

In the frame rotating at the drive frequency $\omega_{01}$, the quantum Langevin equations (QLEs) describing the system dynamics are given by
\begin{equation}\label{QLEs1}
  \begin{split}
     \dot{a} & =-(i\Delta_{a}+\kappa_{a})a-ig_{1}m_{1}-ig_{2}m_{2}+\sqrt{2\kappa_{a}}a^{in}, \\
      \dot{m}_j & =-(i\Delta_{m_j}+\kappa_{m_j} )m_j-iG_{0j}m_j q_j-ig_ja+\Omega_j+\sqrt{2\kappa_{m_j}}m_j^{in}, \\
       \dot{q}_{j}& =\omega_{b_j}p_{j}, \\
       \dot{p}_j& =-\omega_{b_j}q_{j}-\gamma_{b_j}p_j-G_{0j}m_j^\dagger m_j +\xi_j,
  \end{split}
\end{equation}
where $\Delta_a=\omega_a-\omega_{01}$,  $\Delta_{m_j}=\omega_{m_j}-\omega_{01}$, and $\kappa_{a}$, $\kappa_{m_j}$ and $\gamma_{b_j}$ are the dissipation rates of the cavity, magnon and mechanical modes, respectively. The Rabi frequency $\Omega_j=\Omega_1\delta_{j1}$ ($j=1,2$) implies only one drive field applied on the magnon mode $m_1$. $a^{in}$ and $m_j^{in}$ are the input noise operators affecting the cavity and magnon modes, whose non-zero correlation functions are $\langle a^{in}(t)a^{in\dagger}(t')\rangle=[N_a(\omega_a)+1]\delta(t-t')$, $\langle a^{in\dagger}(t)a^{in}(t')\rangle=N_a(\omega_a)\delta(t-t')$, $\langle m^{in}_j(t)m^{in\dagger}_j(t')\rangle=[N_{m_j}(\omega_{m_j})+1]\delta(t-t')$ and $\langle m^{in\dagger}_j(t)m^{in}_j(t')\rangle=N_{m_j}(\omega_{m_j})\delta(t-t')$. The Langevin force operator $\xi_j$ is accounting for the mechanical Brownian motion, which is autocorrelated as $\langle \xi_j(t)\xi_j(t')+\xi_j(t')\xi_j(t)\rangle\approx \gamma_{b_j}[2N_{b_j}(\omega_{b_j})+1]\delta(t-t')$, where we consider a high quality factor $Q_{b}=\omega_b/\gamma_b \gg 1$ for the mechanical oscillators to validate the Markov approximation~\cite{Markov}. Here,  $N_k(\omega_k)=\left[{\rm exp}(\frac{\hbar\omega_k}{k_B T})-1\right]^{-1}$ ($k=a,m_j,b_j$) are the equilibrium mean thermal photon, magnon, and phonon numbers, respectively, at the environmental temperature $T$, with $k_B$ being the Boltzmann constant.

The strong drive field leads to large amplitudes of the magnon modes and cavity mode due to the magnon-cavity coupling, $|\langle m_j \rangle|, |\langle a\rangle| \gg 1$. This allows us to linearize the nonlinear QLEs~\eqref{QLEs1} by writing each mode operator as a classical average plus a fluctuation operator with zero mean value, i.e., $O=\langle O\rangle+\delta O$ ($O=a, m_j, q_j, p_j$), and by neglecting small second-order fluctuation terms. Substituting the above mode operators into Eq.~\eqref{QLEs1} yields two sets of linearized Langevin equations, respectively, for classical averages and fluctuation operators. By solving the former set of equations in the time scale where the system evolves into a stationary state, we obtain the solution of the steady-state averages, which are $\langle p_j\rangle=0$, $\langle q_j\rangle = -G_{0j}|\langle m_j\rangle|^2/\omega_{b_j}$, $\langle a\rangle = -i \big(g_{1}\langle m_1\rangle+g_2\langle m_2\rangle \big) / (i\Delta_a+\kappa_a)$, and 
\begin{equation}\label{mj}
  \begin{split}
     \langle m_1\rangle & =\frac{\Omega_1(i\Delta_a+\kappa_a)}{g_1^2+(i\tilde{\Delta}_{m_1}+\kappa_{m_1})(i\Delta_a+\kappa_a)-\frac{g_1^2 g^2_2}{g_2^2+(i\tilde{\Delta}_{m_2}+\kappa_{m_2})(i\Delta_a+\kappa_a)}} , \\
     \langle m_2\rangle & =-\frac{\Omega_1(i\Delta_a+\kappa_a)\frac{g_1g_2}{g_1^2+(i\tilde{\Delta}_{m_1}+\kappa_{m_1})(i\Delta_a+\kappa_a)}}{g_2^2+(i\tilde{\Delta}_{m_2}+\kappa_{m_2})(i\Delta_a+\kappa_a)-\frac{g_1^2 g^2_2}{g_1^2+(i\tilde{\Delta}_{m_1}+\kappa_{m_1})(i\Delta_a+\kappa_a)}},
  \end{split}
\end{equation}
where $\tilde{\Delta}_{m_j}=\Delta_{m_j}+G_{0j}\langle q_j\rangle$ is the effective magnon-drive detuning, which includes the magnetostriction induced frequency shift. Typically, this frequency shift is negligible (because of a small $G_0$~\cite{Tang16,Davis,Jie22}) with respect to the optimal detuning used in this work, i.e., $|\tilde{\Delta}_{m_j} - \Delta_{m_j}| \ll |\Delta_{m_j}| \approx \omega_{b_j}$. Therefore, in what follows we can safely assume $\tilde{\Delta}_{m_j} \approx \Delta_{m_j}$.

The set of the linearized QLEs for the system fluctuations can be written in the matrix form
\begin{equation}\label{lLQEs}
  \dot{u}(t)=Au(t)+n(t),
\end{equation}
where $u(t)=\big[\delta X(t),\delta Y(t),\delta x_1(t),\delta y_1(t),\delta x_2(t),\delta y_2(t),\delta q_1(t),$  $\delta p_1(t),\delta q_2(t),\delta p_2(t) \big]^T$ is the vector of the quadrature fluctuations, and $n(t)=\Big[ \sqrt{2\kappa_a} \, X^{in}(t), \sqrt{2\kappa_a} \, Y^{in}(t), \sqrt{2\kappa_{m_1}} \, x_1^{in}(t),$ $ \sqrt{2\kappa_{m_1}} \, y_1^{in}(t), \, \sqrt{2\kappa_{m_2}} \, x_2^{in}(t), \, \sqrt{2\kappa_{m_2}} \, y_2^{in}(t), \,\,\, 0, \,\,\, \xi_1(t), \,\,\, 0, \,\,\, \xi_2(t)\Big]^T$ is the vector of the input noises entering the system, with $\delta X\,\,\,{=}\,\,\left(\delta a + \delta a^\dagger \right)/\sqrt{2}$, $\delta Y \,\,\, {=} \,\,\, i \, \left(\delta a^\dagger-\delta a \right)/\sqrt{2}$, $\delta x_j=\left (\delta m_j + \delta m_j^\dagger\right )/\sqrt{2}$, and $\delta y_j=i \left(\delta m_j^\dagger-\delta m_j \right)/\sqrt{2}$, and the drift matrix $A$ is given by
\begin{widetext}
\begin{equation}\label{matrixA}
  A=\begin{pmatrix}
      -\kappa_a & \Delta_a & 0 & g_1 & 0 & g_2 & 0 & 0 & 0 & 0 \\
      -\Delta_a & -\kappa_a & -g_1 & 0 & -g_2 & 0 & 0 & 0 & 0 & 0 \\
      0 & g_1 & -k_{m_1} &{\Delta}_{m_1} & 0 & 0 & -{\rm Re}\left[G_1\right] & 0 & 0 & 0 \\
      -g_1 & 0 & -{\Delta}_{m_1} & -\kappa_{m_1} & 0 & 0 & -{\rm Im}\left[G_1\right] & 0 & 0 & 0 \\
      0 & g_2 & 0 & 0 & -\kappa_{m_2} &{\Delta}_{m_2} & 0 & 0 & -{\rm Re}\left[G_2\right] & 0 \\
      -g_2 & 0 & 0 & 0 & -{\Delta}_{m_2} & -\kappa_{m_2} & 0 & 0 & -{\rm Im}\left[G_2\right] & 0 \\
      0 & 0 & 0 & 0 & 0 & 0 & 0 & \omega_{b_1} & 0 & 0 \\
      0 & 0 & -{\rm Im}\left[G_1\right] & {\rm Re} \left[G_1\right] & 0 & 0 & -\omega_{b_1} & -\gamma_{b_1} & 0 & 0 \\
      0 & 0 & 0 & 0 & 0 & 0 & 0 & 0 & 0 & \omega_{b_2} \\
      0 & 0 & 0 & 0 & -{\rm Im}\left[G_2\right] & {\rm Re}\left[G_2\right] & 0 & 0 & -\omega_{b_2} & -\gamma_{b_2}
    \end{pmatrix},
\end{equation}
\end{widetext}
with $G_j=i\sqrt{2}G_{0j}\langle m_j\rangle$ being the effective magnomechanical coupling rate, which is generally complex.

Because of the linearized dynamics of the system and the Gaussian nature of the input noises, the steady state of the system's quantum fluctuations is a five-mode Gaussian state. The state can be completely described by a $10\times10$ covariance matrix (CM) $\mathcal{V}$ with entries defined as $\mathcal{V}_{ij} = \langle u_{i}(t) u_{j}(t') + u_{j} (t') u_{i} (t) \rangle /2$ $(i,j = 1,2,\ldots,10)$. The CM $\mathcal{V}$ can be obtained by directly solving the Lyapunov equation
\begin{equation}\label{lyap}
  A\mathcal{V}+\mathcal{V}A^{T}=-D,
\end{equation}
where the diffusion matrix $D$ is defined as $D_{ij} = \langle n_i(t)n_j(t')\,{+}\,n_j(t')n_i(t)\rangle/\big[ 2\delta(t{-}t') \big]$ and given by $D={\rm diag}[\kappa_{a}(2N_{a} \,\, {+} \,\,1), \kappa_{a}(2N_{a}+1), \kappa_{m_1}(2N_{m_1}+1), \kappa_{m_1}(2N_{m_1}\, {+} \,\,1),$ $\kappa_{m_2}(2N_{m_2}+1), \kappa_{m_2}(2N_{m_2}+1), 0, \gamma_{b_1}(2N_{b_1}+1), 0, \gamma_{b_2}(2N_{b_2}+1)]$.

When the CM of the system fluctuations is obtained, one can then extract the state of the interesting modes by removing in $\mathcal{V}$ the rows and columns related to the uninteresting modes and study their entanglement properties. We use the logarithmic negativity~\cite{LN} to quantify the Gaussian bipartite entanglement, of which the definition is provided in the Appendix.

\begin{figure}[t]
\includegraphics[width=\linewidth]{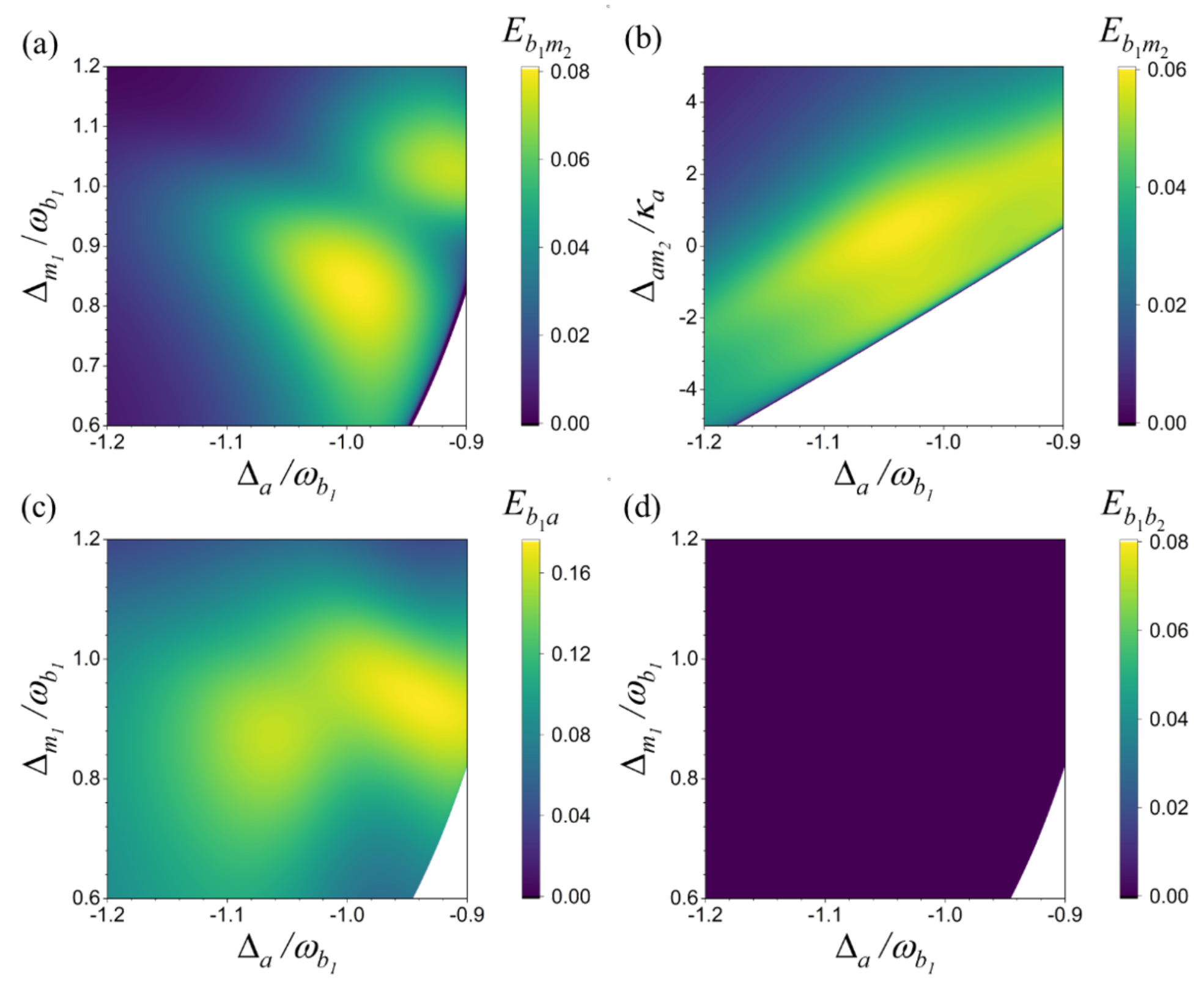}
\caption{Nonlocal phonon-magnon entanglement $E_{b_1m_2}$ versus (a) $\Delta_a$ and ${\Delta}_{m_1}$, and (b) $\Delta_a$ and $\Delta_{am_2} \equiv \omega_a - \omega_{m_2}$. (c) The phonon-cavity entanglement $E_{b_1a}$ and (d) the mechanical entanglement $E_{b_1b_2}$ versus $\Delta_a$ and ${\Delta}_{m_1}$. In (d) the whole purple region denotes $E_{b_1b_2}=0$. We take $\Delta_{am_2}=0.9 \kappa_a$ in (a), (c) and (d), and ${\Delta}_{m_1}=0.95 \omega_{b_1}$ in (b). The blank areas denote the unstable region where the stability condition is not fulfilled. See the text for other parameters.}\label{fig3}
\end{figure}

In Fig.~\ref{fig3}, we show relevant steady-state bipartite entanglements generated in the first step. The blank areas are the parameter regimes in which the system is unstable, reflected by the fact that the drift matrix $A$ has at least one positive eigenvalue. We have employed the following parameters~\cite{Tang16,Davis,Jie22,bridge,Fan2}: $\omega_a/2\pi=10$ GHz, $\omega_{b_1}/2\pi=17$ MHz, $\omega_{b_2}/2\pi=12$ MHz, $\gamma_{b}/2\pi=100$ Hz, $\kappa_a/2\pi=1$ MHz, $\kappa_{m}=\kappa_a$, $g_1/2\pi=5$ MHz, $g_2/2\pi=1$ MHz, $G_{01}/2\pi \simeq G_{02}/2\pi=10$ Hz, and $T=10$ mK. The amplitude of the drive magnetic field $B_{1}=\SI{4.8e-4}{T}$, corresponding to the drive power $P_1 \simeq 1.1$ mW for two YIG micro bridges (approximated as two cuboids) with dimensions of 13.7 $\times$ 3 $\times$ 1 $\mu$m$^3$ and 16.4 $\times$ 3 $\times$ 1 $\mu$m$^3$. This milliwatt power might lead to heating induced temperature rise. To avoid the possible heating effect, one should improve the bare coupling $G_{0j}$ to reduce the power.
Since the mechanical resonance frequency mainly depends on the length and thickness of the beam~\cite{bridge}, the sizes of the beams are chosen corresponding to the mechanical frequencies we adopted. To determine the power, we have used the relation between the drive magnetic field $B_1$ and the power $P_1$, i.e., $B_1=\sqrt{2\mu_0 P_1/(lwc)}$~\cite{Fan2}, with $\mu_0$ being the vacuum magnetic permeability, $c$ the speed of the electromagnetic wave propagating in vacuum, and $l$ and $w$ the length and width of the YIG cuboid.
The nonlocal phonon-magnon entanglement $E_{b_1m_2}$ (Fig.~\ref{fig3}(a)-(b)) is the result of the phonon-cavity entanglement $E_{b_1a}$ (Fig.~\ref{fig3}(c)) and the cavity-magnon ($a$-$m_2$) beamsplitter (state-swap) interaction. The phonon-cavity entanglement $E_{b_1a}$ is obtained by using the results of Ref.~\cite{Jie18}, which studies a tripartite magnon-photon-phonon system (i.e., the $m_1$-$a$-$b_1$ subsystem here). As shown in Ref.~\cite{Jie18}, a red-detuned {\it strong} drive field on the magnon mode $m_1$ (with $\Delta_{m_1} \approx \omega_{b_1}$) cools the hot mechanical mode $b_1$ (a precondition for preparing quantum states) and, simultaneously, activates the magnomechanical PDC interaction, which produces the magnomechanical entanglement $E_{b_1m_1}$. The latter process occurs only when the pump field is sufficiently strong, such that the weak coupling condition $|G_1| \ll \omega_{b_1}$ for taking the rotating-wave (RW) approximation to obtain the cooling interaction $\propto m_1^\dag b_1 + m_1 b_1^\dag$ is no longer satisfied, and the counter-RW terms $\propto m_1^\dag b_1^\dag + m_1 b_1$ (responsible for the PDC interaction) start to play the role. The magnomechanical entanglement is partially transferred to the phonon-cavity subsystem with $E_{b_1a}>0$ (Fig.~\ref{fig3}(c)) when ${\Delta}_a \approx -\omega_{b_1}$~\cite{Jie18}. The optimal detunings $\Delta_{m_1} \approx -\Delta_a \approx \omega_{b_1}$ imply that the cavity and the magnon mode $m_1$ are respectively resonant with the two mechanical sidebands of the drive field, see Fig.~\ref{fig2}. 

Since the cavity-magnon ($m_2$) state-swap operation works optimally when the two modes are resonant. The entanglement $E_{b_1m_2}$ is maximized for a nearly zero cavity-magnon detuning $\Delta_{am_2} \approx 0$, as confirmed by Fig.~\ref{fig3}(b). The detuning $\Delta_{am_2}$ up to several cavity linewidth will significantly hinder the transfer of the entanglement. The fact that the entanglement $E_{b_1m_2}$ is transferred from the entanglement  $E_{b_1a}$ can also be seen from the complementary distribution of the entanglement in Figs.~\ref{fig3}(a) and \ref{fig3}(c), indicating the entanglement flow among the subsystems switched on by the beamsplitter coupling.

It is worth noting that in the first step the two mechanical modes are not entangled, as confirmed by $E_{b_1b2}=0$ in Fig.~\ref{fig3}(d) in a wide range of parameters. This is mainly because the magnon mode $m_2$ is driven by the {\it nearly resonant} cavity field, which is not the condition for cooling the mechanical mode $b_2$, or realizing the state-swap interaction between the two modes $m_2$ and $b_2$. Instead, a red-detuned microwave field should be used to drive the magnon mode $m_2$ to realize the state-swap operation $m_2 \leftrightarrow b_2$, such that the phonon-magnon entanglement $E_{b_1m_2}$ can be transferred to the two mechanical modes, as will be discussed in the next section.

\section{Entanglement between two mechanical modes}
\label{2step}

\begin{figure}[t]
\hskip-0.0cm \includegraphics[width=\linewidth]{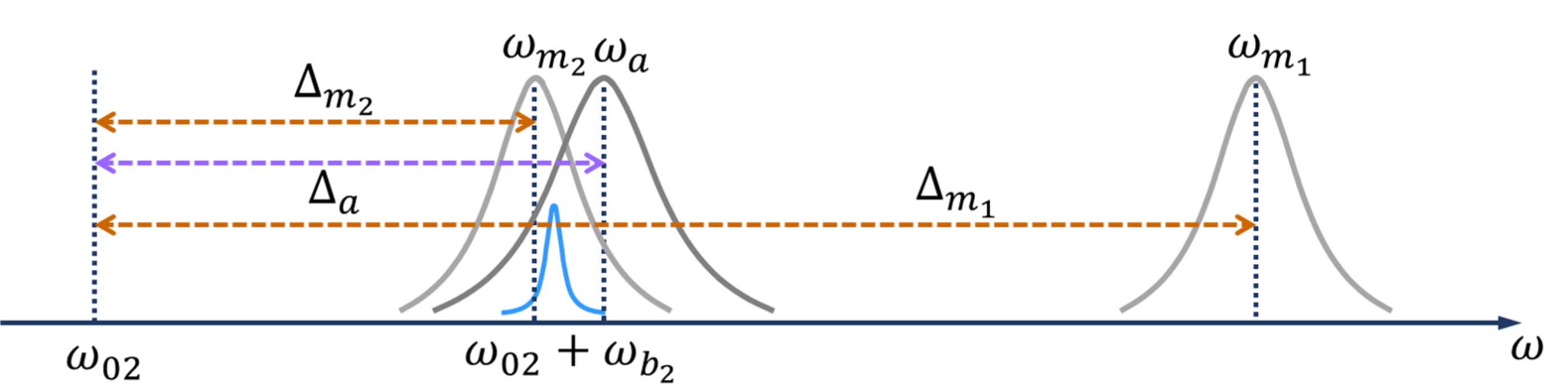}
\caption{Mode and drive frequencies used in the second step. A red-detuned pulsed drive with frequency $\omega_{02}$ is used to activate the magnomechanical state-swap operation $m_2 \leftrightarrow b_2$, which transfers the phonon-magnon entanglement $E_{b_1 m_2}$ generated in the first step to the two mechanical modes. }
\label{fig4}
\end{figure}

In this section, we show how to transfer the phonon-magnon entanglement $E_{b_1m_2}$ prepared in the first step to the two mechanical modes. To this end, we apply a red-detuned {\it pulsed} microwave drive on the magnon mode $m_2$ (see Fig.~\ref{fig4}) to activate the local magnomechanical state-swap interaction $m_2 \leftrightarrow b_2$. The pulsed drive should be fast as the entanglement $E_{b_1m_2}$ will quickly decay as soon as the continuous drive is turned off in the first step. To simplify the model, we use a flattop microwave pulse to drive the magnon mode and then the model and the treatment used in Sec.~\ref{1step} for a continuous drive are still valid for the case of a flattop pulse drive. Differently, we shall solve the dynamical solutions rather than the steady-state solutions. 

The QLEs remain the same as in Eq.~\eqref{QLEs1}, except that all the detunings are redefined with respect to the frequency $\omega_{02}$ of the pulsed drive and the Rabi frequency is associated with the pulsed drive, i.e., $\Delta_a=\omega_a-\omega_{02}$, $\Delta_{m_j}=\omega_{m_j}-\omega_{02}$, and $\Omega_j=\Omega_2\delta_{j2}$. The dynamics of the quantum fluctuations of the system can still be described by Eq.~\eqref{lLQEs} but with dynamical $\langle m_j\rangle(t)$ and thus $G_j(t)$ in the drift matrix, due to a pulsed drive. 



The dynamical CM $\mathcal{V}(t)$ can be obtained by~\cite{Jie17}
\begin{equation}\label{Vint}
  \mathcal{V}(t)=M(t) \mathcal{V}_0 M(t)^T+\int_{0}^{t}ds\, M(s) D M(s)^T,
\end{equation}
where $t$ is the duration of the pulsed drive, $M(t)=e^{\int_{0}^{t} A(\tau) d\tau}$, and $\mathcal{V}_0$ is the CM of the initial state of the system when simultaneously turn on (off) the drive on the magnon mode $m_2$ ($m_1$), which is obtained in the first step by solving the Lyapunov equation~\eqref{lyap}. When the dynamical CM is achieved, we can then study the dynamics of the entanglement.

\begin{figure}[t]
\includegraphics[width=\linewidth]{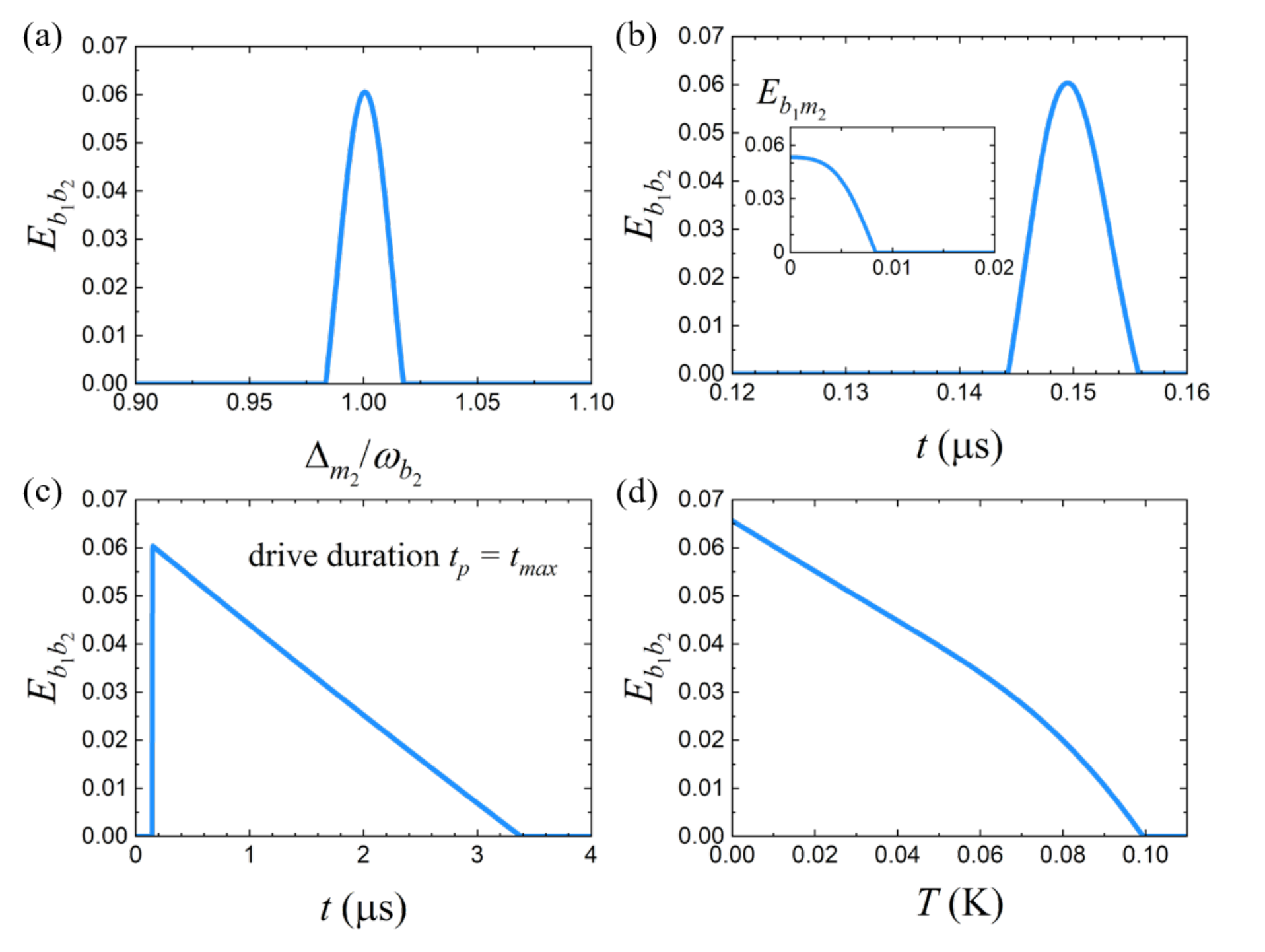}
\caption{Mechanical entanglement $E_{b_1b_2}$ versus (a) detuning $\Delta_{m_2}$ at the optimal pulse duration $t_{\rm max}$; (b) pulse duration $t$ at the optimal detuning $\Delta_{m_2}=\omega_{b_2}$. The inset shows that the phonon-magnon entanglement $E_{b_1m_2}$ dies out around $t \simeq 0.01$ $\mu$s. (c) $E_{b_1b_2}$ versus time when the drive is turned off at $t_{\rm max}$. Soon after $t_{\rm max}$, the two mechanical oscillators evolve freely with the only coupling to their local thermal baths. (d) $E_{b_1b_2}$ versus bath temperature at the optimal detuning and pulse duration. We take $\Delta_a = -0.95\omega_{b_1}$, $\Delta_{m_1}=0.95\omega_{b_1}$ (as defined in the first step), and $\Delta_{am_2}=0.9\kappa_a$ in all plots. The other parameters are the same as in Fig.~\ref{fig3}. }
\label{fig5}
\end{figure}

Figure~\ref{fig5}(a) shows the mechanical entanglement $E_{b_1b_2}$ is maximized at the detuning $\Delta_{m_2} \approx \omega_{b_2}$, which is the optimal detuning for realizing the magnomechanical state-swap interaction, by which the previously generated phonon-magnon entanglement $E_{b_1m_2}$ is transferred to the mechanical modes. Note that Fig.~\ref{fig5}(a) is plotted at the optimal pulse duration $t_{\rm max}$ for a given detuning, yielding a maximal $E_{b_1b_2}$.  Figure~\ref{fig5}(b) shows $E_{b_1b_2}$ versus the pulse duration $t$ for the optimal detuning. As is shown, there is a time window for the presence of the entanglement. The mechanical entanglement emerges a short while after the phonon-magnon entanglement dies out (c.f. the inset of Fig.~\ref{fig5}(b)).
When $E_{b_1b_2}$  reaching its maximum at $t_{\rm max}$, we then turn off the pulse drive to decouple the mechanics from the rest of the system to protect the entanglement. The two mechanical oscillators then evolve almost freely, and their entanglement is affected by their local thermal baths, which lasts for a much longer time due to a small mechanical damping and a low bath temperature. This is clearly shown in Fig.~\ref{fig5}(c) (c.f. Fig.~\ref{fig5}(b)). The mechanical entanglement is robust against bath temperature and the maximal entanglement survives up to $T \simeq 100$ mK, as illustrated in Fig.~\ref{fig5}(d). We have used a drive power $P_2 = 1.3$ mW, giving the amplitude of the drive magnetic field $B_2=4.8 \times 10^{-4}$ T.

Lastly, we discuss how to detect the mechanical entanglement. The entanglement can be verified by measuring the CM of the two mechanical modes. The mechanical quadratures can be measured by coupling the deformation displacement to an optical cavity that is driven by a weak red-detuned light to transfer the mechanical state to the optical field.  By homodyning two cavity output fields one can then obtain the CM of the mechanical modes~\cite{John,Lehnert13}.

\section{Conclusion and discussion}
\label{conc}

We present a protocol to entangle two mechanical vibration modes in a cavity magnomechanical system. The protocol contains two steps by applying successively two drive fields on two magnon modes to activate different functions of the nonlinear magnetostrictive interaction, namely, the magnomechanical PDC and state-swap operations. We show that the entanglement between two mechanical vibration modes of two YIG crystals can be achieved by fully exploiting the above magnetostrictive functions and the cavity-magnon state-swap interaction. We remark that our protocol is valid for any magnomechanical system of ferrimagnets or ferromagnets, spherical~\cite{Tang16,Davis,Jie22} or nonspherical structures~\cite{bridge,Fan2}, as long as they possess the dispersive coupling between magnons and phonons. The work may find important applications in many studies that require the preparation of macroscopic entangled states.

\section*{Acknowledgments}

This work has been supported by National Key Research and Development Program of China (Grant No. 2022YFA1405200) and National Natural Science Foundation of China (Nos. 92265202 and 11874249).

\section*{Appendix}

The logarithmic negativity is used to quantify the Gaussian bipartite entanglement, which is defined as
\begin{equation}
E_N\equiv \rm{max} \big[0,-\rm{ln}\,2\tilde{\nu}_- \big],
\end{equation}
where $\tilde{\nu}_{-}=\rm{min}\,eig |i\Omega_2\tilde{\mathcal{V}}_4|$ ($\Omega_2=\oplus^2_{j=1}i\sigma_y$ and $\sigma_y$ is the $y$-Pauli matrix) is the minimum symplectic eigenvalue of the CM $\tilde{\mathcal{V}}_4 = \mathcal{P}\mathcal{V}_4 \mathcal{P}$, with $\mathcal{V}_4$ being the $4\times4$ CM of two relevant modes, obtained by removing in $\mathcal{V}$ the rows and columns of the uninteresting modes, and $\mathcal{P}= \rm{diag}(1,-1,1,1)$ being the matrix that implements the partial transposition of the CM.

\end{document}